\documentclass[%
 preprint,
 superscriptaddress,
preprintnumbers,
nofootinbib,
 amsmath,amssymb,
11pt
]{revtex4-1}
\pdfoutput=1
\usepackage{amsmath}
\usepackage{amssymb}
\usepackage{slashed}
\usepackage{graphicx}
\usepackage{graphics}
\usepackage{epsfig}
\usepackage{color}
\usepackage{xcolor}
\usepackage{soul}
\usepackage{hyperref}
\usepackage[section]{placeins}
\usepackage{mathrsfs}
\usepackage{dsfont}
\usepackage{comment}
\usepackage{mathtools}

\newcommand{\be}{\begin{equation}}
\newcommand{\ee}{\end{equation}}
\newcommand{\bea}{\begin{eqnarray}}
\newcommand{\eea}{\end{eqnarray}}

\begin{document}
\title{Towards a bound on the Higgs mass in causal set quantum gravity
}

\author{Gustavo P.~de Brito}
   \email{gustavo@cp3.sdu.dk}
   \affiliation{CP3-Origins, University of Southern Denmark, Campusvej 55, DK-5230 Odense M, Denmark}

\author{Astrid Eichhorn}
   \email{eichhorn@cp3.sdu.dk}
   \affiliation{CP3-Origins, University of Southern Denmark, Campusvej 55, DK-5230 Odense M, Denmark}

\author{Ludivine Fausten}
   \affiliation{Fields and Strings Laboratory, Institute of Physics, \`Ecole Polytechnique Fédérale de Lausanne (EPFL), Rte de la Sorge, BSP 728, CH-1015 Lausanne, Switzerland}
\affiliation{CP3-Origins, University of Southern Denmark, Campusvej 55, DK-5230 Odense M, Denmark}

\begin{abstract} 
In the Standard Model of particle physics, the mass of the Higgs particle can be linked to the scale at which the Standard Model breaks down due to a Landau pole/triviality problem: for a Higgs mass somewhat higher than the measured value, the Standard Model breaks down before the Planck scale. We take a first step towards investigating this relation in the context of causal set quantum gravity. We use a scalar-field propagator that carries the imprints of spacetime discreteness in a modified ultraviolet behavior that depends on a nonlocality scale. We investigate whether the modification can shift the scale of the Landau pole in a scalar field theory with quartic interaction. We discover that the modifications speed up the onset of the Landau pole considerably, so that the scale of new physics occurs roughly at the nonlocality scale. Our results call into question, whether a separation between the nonlocality scale and the discreteness scale, which is postulated within causal set quantum gravity, and which has been argued to give rise to phenomenological consequences, is in fact achievable.\\
Methodologically, our paper is the first to apply continuum functional Renormalization Group techniques in the context of a causal-set inspired setting.
\end{abstract}

\maketitle
\section{Introduction and motivation}
Is causal set quantum gravity a viable quantum theory of gravity? We approach this question from a phenomenological perspective. We follow the ``matter matters" paradigm, which says that observational tests  of quantum gravity can be derived from the interplay of quantum gravity with matter \cite{Eichhorn:2022gku}.
The paradigm does not contradict the standard viewpoint that quantum gravity is dynamically irrelevant at particle physics scales, such that, e.g., at the LHC, gravity-mediated scattering is completely negligible. Instead, the paradigm states that fundamental parameters of particle physics (e.g., quantum numbers and coupling values) are (partially) constrained by the interplay with quantum gravity at quantum-gravity scales.

Our starting point is the causal-set d'Alembertian \cite{Sorkin:2007qi,Benincasa:2010ac,Dowker:2013vba,Belenchia:2015hca}, which gives rise to a scalar propagator. As a consequence of causal-set discreteness, the propagator is modified in the ultraviolet (UV) \cite{Aslanbeigi:2014zva,Belenchia:2014fda}.
In \cite{Aslanbeigi:2014zva}, the idea has been put forward that the modified behavior of the propagator in the UV could serve as a UV regulator and improve the UV behavior of QFTs. Here, we explore this idea. The crucial point is not whether or not an individual loop diagram diverges -- the theory of renormalization allows us to understand and control these divergences. Instead, the crucial point is the scale dependence of the (renormalized) coupling as a function of the Renormalization Group (RG) scale: If this scale dependence leads to a Landau pole, the theory is expected to break down at this scale. \footnote{More precisely, it is perturbation theory that breaks down at a Landau pole. Nonperturbative techniques are required to determine whether the theory continues to be viable in a strongly coupled regime. Within $\phi^4$ theory in four dimensions, lattice studies have shown that the theory does indeed break down in the UV: if the UV cutoff of the theory is pushed to infinity, the theory becomes noninteracting, i.e., trivial, in the IR.} To investigate this, we calculate the beta function of the quartic scalar interaction, using the causal-set inspired propagator. Within standard quantum field theory, this interaction runs into a Landau pole in the UV, which signals the incompleteness of $\lambda\, \phi^4$ theory and the need for new physics. Given a value of $\lambda$ at low energies, one can calculate the scale of the Landau pole. We will investigate how the scale of the Landau pole $k_{\rm Landau}$ is related to the nonlocality scale $k_{\rho}=\rho^{1/4}$.

For this investigation, we need a tool that allows us to account for the effect of quantum fluctuations of the scalar field and to derive the beta function based on the modified propagator. The functional Renormalization Group (RG) is a suitable tool, given the easiness with which various modified (and/or dressed) propagators can be put in and their effect on beta functions explored.

This paper is structured as follows: Because continuum functional RG techniques have not yet been used in causal set quantum gravity\footnote{A discrete version of the functional RG, developed in \cite{Eichhorn:2013isa}, has been adapted to causal sets in \cite{Eichhorn:2017bwe}.}, we attempt at make our paper accessible both to the causal-set community as well as to practitioners of the functional RG. We thus introduce the salient features of causal set quantum gravity in Sec.~\ref{sec:CSQG} and review the scalar d'Alembertian and the resulting momentum-space representation of the propagator  on discrete Minkowski spacetime.  We also define a causal-set inspired, Euclidean counterpart of the propagator that forms the basis of our calculation. In Sec.~\ref{sec:FRG} we introduce the functional RG and discuss the distinct scales involved in our setting and how the functional RG enables us to calculate the effect of quantum fluctuations at the different scales. Because we consider it important to clarify exactly which assumptions underlie any connection of a quantum-gravity approach to phenomenology, we dedicate Sec.~\ref{sec:assumptions} to spelling out our assumptions in detail. In Sec.~\ref{sec:results} we discuss our results. First, we review the triviality bound on the Higgs mass in standard QFT, in order to then discuss where our results deviate due to the nonlocality scale. In Sec.~\ref{sec:conclusions} we provide conclusions as well as two outlooks, one on physical implications and the broader context of our work and one on future methodological developments that are possible based on the present work.

\section{Causal set quantum gravity}\label{sec:CSQG}
To set the stage, we review the key features of causal set quantum gravity and the scalar field propagator on a causal set background.
\subsection{Review of causal set quantum gravity}
Causal set quantum gravity \cite{Bombelli:1987aa}, see \cite{Surya:2019ndm} for a recent review, postulates that spacetime is fundamentally discrete. Spacetime manifolds and associated structures, such as tangent spaces, continuum fields and also the spacetime metric, are regarded as approximative structures that provide a good description of physics at spacetime distances much larger than the discreteness scale. At the fundamental level, only spacetime points and causal relations remain of the continuum spacetime geometry. \\
More specifically, a  causal set $C$ is a collection of spacetime points $x_i \in C$, with a causal relation $\preceq$, such that $x_i \preceq x_j$, if $x_i$ is in the causal past of $x_j$.\footnote{It holds that $x_i \preceq x_i$.} The causal relation satisfies transitivity, as it intuitively should,
\be
x_i \preceq x_j \, \mbox{ and }\, x_j \preceq x_k\,\, \Rightarrow x_i \preceq x_k,\, \, \forall x_{i,j,k} \in C .
\ee
Causal set quantum gravity excludes spacetimes with closed timelike curves\footnote{There is therefore no causal set that corresponds to the maximum analytic extension of the Kerr spacetime; it is, however, an open question what the structure of spinning black holes is in causal sets.}, i.e.,
\be
x_i \preceq x_j\, \mbox{ and }\, x_j \preceq x_i \,\,\Rightarrow x_i = x_j\, \, \, \forall x_{i,j} \in C.
\ee
Both of these properties hold for spacetime points on a continuum manifold that does not contain closed timelike curves. The third property satisfied by a causal set is what makes it distinct from a continuum manifold by imposing spacetime discreteness:
\be
|\{x_k \in C:\, x_i \preceq x_k \preceq x_j\}|< \infty \,,
\ee
 where $|\{\cdots\}|$ represents the cardinality of a set. 
This last property is called local finiteness and says that there is only a finite number of spacetime points in the causal interval between any two spacetime points.

Starting from a given causal set, the construction of an associated continuum manifold\footnote{There is no one-to-one correspondence, because all continuum manifolds which differ from each other only at spacetime scales smaller than the discreteness scale are equivalent in the sense that they can be associated to the same causal set.} is a challenging and in general unsolved problem. Progress towards a solution has been made, in that topological and geometric quantities have been constructed from the causal set, which characterize an associated continuum geometry. These quantities include a characterization of the spacetime topology \cite{Major:2006hv,Major:2009cw}, spacetime dimensionality \cite{Myrheim:1978ce,Reid:2002sj,Eichhorn:2013ova,Glaser:2013pca,Roy:2013}, geodesic distance \cite{Brightwell:1990ha,Rideout:2008rk,Eichhorn:2018doy}, Ricci curvature \cite{Benincasa:2010ac} and higher-order curvature invariants \cite{deBrito:2023axj}, see also the review \cite{Surya:2019ndm}.

Starting from a continuum manifold, an associated causal set can be obtained via a Poisson sprinkling, i.e., a random selection of spacetime points, with the number of spacetime points per volume following from a Poisson distribution at sprinkling density $\rho_{\ell} = 1/\ell^4$, where $\ell$ is the resulting discreteness scale, typically assumed to be the Planck scale. The causal relation $\preceq$ between the selected spacetime points follows from the spacetime metric; after the causal relations are derived, the spacetime manifold and associated metric can be discarded, and a causal set remains. The randomness of the distribution is key to avoid the selection of a preferred frame and the associated breaking of Lorentz invariance that any non-random distribution (e.g., a regular spacetime lattice) implies, see \cite{Bombelli:2006nm}. As an important consequence, causal sets encoding an infinite spacetime volume are generically graphs with infinite valency and for a finite-size causal set the valency generically increases with the spacetime volume (although spacetime curvature may limit this growth). 

The discrete-to-continuum correspondence between a causal set and a Lorentzian manifold is based on the Hawking-Malament theorem \cite{Hawking:1976fe,Malament:1977} which says that in a past-and- future-distinguishing manifold, the causal relations encode the full conformal metric. The conformal factor can also be recovered from a causal set: because it is discrete, the number of spacetime points in a causal set (or a subset of it) correspond to the spacetime volume, with one spacetime point on average corresponding to a volume $1/\rho_{\ell}$. Thus, the discrete-to-continuum correspondence can be stated as ``order + number = geometry".

Our focus will be on such causal sets that are sprinklings into four-dimensional Minkowski spacetime, because on those, the discrete counterpart of a d'Alembertian has been constructed.

\subsection{Review of the scalar propagator on a causal set}
In \cite{Sorkin:2007qi,Benincasa:2010ac,Dowker:2013vba,Belenchia:2015hca}, a discrete d'Alembertian acting on a scalar field has been put forward. It follows the usual intuition that the discrete counterpart of a derivative operator is built out of alternating sums of the field evaluated at neighboring points. For causal sets, these neighbors are the causally related points. For the present purposes, we do not need the discrete form of the d'Alembertian on a single causal set; instead we are interested in its expectation value averaged over sprinklings into $3+1$ dimensional Minkowski spacetime.\footnote{This may account for part of the path integral over causal sets: if an action can be found that suppresses non-manifoldlike causal sets through destructive interference (see \cite{Benincasa:2010ac,Loomis:2017jhn,Carlip:2022nsv} for progress in this direction) and that results in constructive interference of causal sets with small enough curvature, then the average over sprinklings may provide a good estimate of the result of performing the path integral over causal sets.} Given a scalar field $\phi(x)$, the expectation value of the discrete d'Alembertian acting on $\phi$, averaged over Poisson sprinklings into $3+1$ dimensional Minkowski spacetime, is given by
\be
\langle\Box_{\rm CS}\phi(x)\rangle= \frac{1}{\ell^2} \bigg(  -\frac{4}{\sqrt{6}} \, \phi(x)  + \frac{4}{\sqrt{6}\ell^4}\, \sum_{i=1}^{4}  C_i  
	\!\int_{J^-(x)}d^4 y \,\sqrt{-g(y)} \,\, \frac{\big(\ell^{-4} \,\mathcal{V}(x,y) \big)^{i-1} }{(i-1)!}e^{\ell^{-4} \, \mathcal{V}(x,y)} 
	\, \phi(y) \bigg)\,,\label{eq:Box}
\ee
with $C_1=1$, $C_2=-9$, $C_3=16$ and $C_4=-8$. $\mathcal{V}(x,y)$ denotes the causal volume inbetween the two points $x$ and $y$ and $J^-(x)$ denotes the causal past of $x$. For details on this expression and its derivation, we refer the reader to \cite{Sorkin:2007qi,Benincasa:2010ac,Dowker:2013vba,Belenchia:2015hca}. In the continuum limit, $\ell \rightarrow 0$, this expression reduces to the continuum d'Alembertian, see \cite{Belenchia:2015hca, Christopherthesis} for details. However, the fluctuations about the expectation value are large and grow with decreasing $\ell$, see \cite{Sorkin:2007qi}. This can be solved by ``smearing" the discrete operator over spacetime distances $\ell_{k}\gg \ell$. The corresponding continuum operator is still of the form Eq.~\eqref{eq:Box}, but with the substitution $\ell \rightarrow \ell_k$ (and correspondingly, $\rho_{\ell} \rightarrow \rho$).

We will work in the corresponding momentum-space representation,  that is obtained by the action of $ \langle \Box_{\rm CS}\rangle$ on a plane wave:
\bea
\langle \Box_{\rm CS} e^{i \, p\cdot x}\rangle&=& f(p^2)e^{i\, p\cdot x},\\
f(p^2)&=& \rho^{1/2} \left( -\frac{4}{\sqrt{6}}+ 4\pi \frac{\rho^{1/4}}{\sqrt{p^2}} 
	\sum_{n=0}^3 \frac{b_n}{n!} \left(\frac{\pi}{24} \right)^n \int_0^{\infty} \!\!ds\, s^{4n+2} \, e^{-\frac{\pi}{24}\,s^4}
	K_1\left(\frac{\sqrt{p^2}}{\rho^{1/4}} \,s\right) \right) \,.
\eea
The coefficients $b_n$ are given by
\be
b_n=\frac{4\,\delta_{n,0} - 36\,\delta_{n,1} + 64\,\delta_{n,2} - 32\,\delta_{n,3}}{\sqrt{6}}.
\ee
$K_1$ is a modified Bessel function of the second kind. Here, we are also using the momentum scale $\rho^{1/4}$, which is related to the nonlocality scale, $\rho^{1/4} = 1/\ell_k$. 

In the IR limit, $p^2 \ll \sqrt{\rho}$, the causal-set effect vanishes and the d'Alembertian reduces to the usual limit
\be
f(p^2) \rightarrow -p^2\,\, \mbox{ for } \frac{p^2}{\sqrt{\rho}}\ll1.
\ee
Modifications from the usual UV behavior arise for non-zero $\rho$:
\be
f(p^2) \rightarrow - \frac{4}{\sqrt{6}}\sqrt{\rho} + \frac{32\pi}{\sqrt{6}}\frac{\rho^{3/2}}{p^4}\,\, \mbox{ for } \frac{p^2}{\sqrt{\rho}}\gg 1.
\ee
The scalar field propagator, i.e., the inverse of $f(p^2)$, behaves just like the usual propagator in the IR; the UV limit is modified and the propagator behaves like
\be
f(p^2)^{-1} \rightarrow - \frac{\sqrt{6}}{4}\frac{1}{\sqrt{\rho}} - \frac{\sqrt{6}}{32\pi}\frac{\rho^{3/2}}{p^4}\label{eq:UVlimit}
\ee
The far UV limit, in which $\frac{p^2}{\sqrt{\rho}} \rightarrow \infty$, is thus characterized by a constant propagator instead of the usual $p^{-2}$ falloff which characterized a local QFT. Thus, the scale $\rho$ can be thought of as a nonlocality scale for the theory. Possible phenomenological consequences of this nonlocality have been analyzed in \cite{Belenchia:2014fda,Belenchia:2015ake,Saravani:2015rva,Belenchia:2016zaa}.

Here, we are not concerned with the consequences for the analytic structure of the QFT, which has been analyzed in \cite{Belenchia:2014fda}. We instead take the propagator as our starting point and analyze its consequences for the triviality problem.

The fact that the usual UV fall-off is absent and the propagator instead features a constant term already implies that quantum fluctuations will be enhanced in the UV. We thus expect that the Landau-pole problem may get worse and that the nonlocality scale $\rho$ will be closely related to the scale at which the quartic scalar field theory breaks down.

\subsection{Euclidean counterpart of the propagator}
To apply functional RG techniques, we require a Euclidean propagator as our starting point. This propagator can in principle be obtained by performing an analytical continuation in the complex-time-plane. Here, we instead do something simpler to \emph{define} a Euclidean counterpart to the Lorentzian propagator, namely a simple substitution $t \rightarrow -i \tau$.  We perform such substitution  without considering the structure of poles and branch cuts of the propagator. The resulting propagator should rather be understood as a definition of a new Euclidean propagator, inspired by its Lorentzian counterpart. As a function of the Euclidean four-momentum $p^2$, we have that
\be
f_{\rm Euclidean} (p^2) = - f(p^2).\label{eq:euclprop}
\ee
In the IR limit, our definition reduces to the standard analytical continuation of the scalar field propagator.

To make it clear that we are using a propagator that is defined in the Euclidean, we call it the \emph{causal-set inspired} propagator.

\section{Functional Renormalization Group techniques}\label{sec:FRG}
The functional RG enables us to integrate out quantum fluctuations according to the value of the Euclidean four-momentum, see \cite{Dupuis:2020fhh} for a recent review. It is therefore well-suited to our task, namely to account for quantum fluctuations of a scalar field on top of a causal set background. As a function of the RG scale $k$ (where $k = \sqrt{k^2}$ with $k^2$ a four-momentum squared), we first integrate out quantum fluctuations whose propagator corresponds to the UV limit of the causal-set propagator. As $k$ is lowered, the propagator transitions to its IR form where it corresponds to the standard scalar propagator in a local QFT.

This is implemented through the suppression of field configurations whose four-momenta squared $p^2$ lie below the RG scale, i.e, which satisfy $p^2<k^2$. This can most conveniently be done by endowing them with a large mass term, because a mass term reduces the propagator and thus suppresses the corresponding quantum field. Accordingly, we add a term $\Delta S_k=\int \frac{d^4p}{(2\pi)^4}\phi(p)R_k(p^2)\phi(-p)$ to the action, where $R_k(p^2)$ is the mass-like regulator term. It satisfies $R_k(p^2)>1$ for $p^2<k^2$ and $R_k(p^2)= 0$ for $p^2>k^2$, such that field configurations with $p^2<k^2$ are suppressed and field configurations with $p^2>k^2$ are unsuppressed. This can for instance be achieved with the choice \cite{Litim:2000ci,Litim:2001up}
\be
R_k(p^2) = (k^2 - p^2) \theta(k^2-p^2),
\ee
with the Heaviside step function $\theta$.

\begin{figure}[!t]
\includegraphics[width=\linewidth,clip=true,trim=0cm 22cm 0cm 1cm]{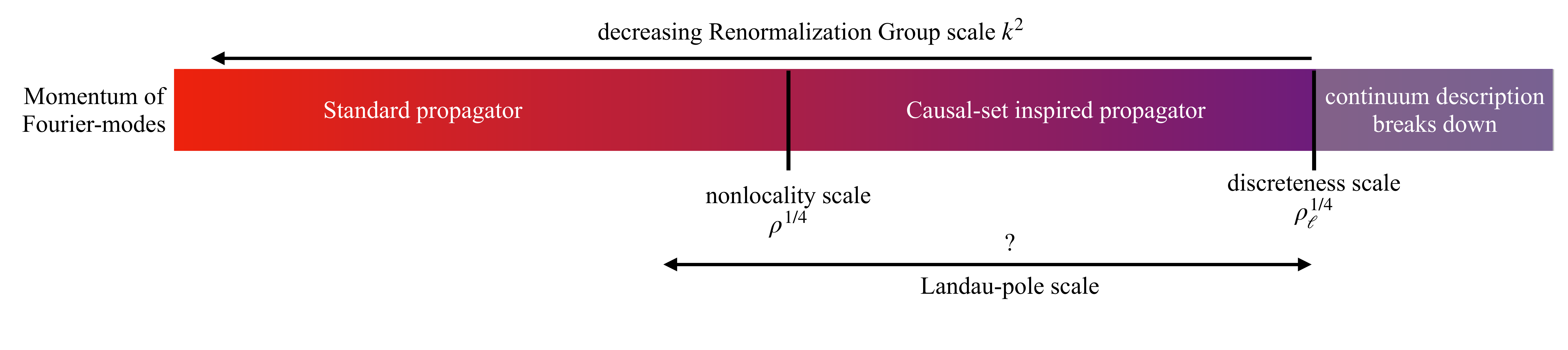}
\caption{\label{fig:illustration}We illustrate the different scales in the problem.}
\end{figure}

To account for the effect of quantum fluctuations, we focus on the effective action $\Gamma$, which is the quantum counterpart to the classical action: its variation gives rise to the equations of motion for the expectation value of the field. Knowing $\Gamma$ is equivalent to having performed the path integral. In the functional RG setup, $\Gamma$ is not calculated at once, but in successive steps, each of which accounts for quantum fluctuations of successively lower momenta. This gives rise to a flowing action $\Gamma_k$, which is obtained from a path integral with $\Delta S_k$ added to the action. We illustrate  the scales involved in the setting in Fig.~\ref{fig:illustration}.

The advantage of working with $\Gamma_k$ is that it satisfies a formally exact functional differential equation \cite{Wetterich:1992yh,Morris:1993qb}, which reads
\be\label{eq:floweq}
k\,\partial_k\Gamma_k = \frac{1}{2} {\rm Tr} \left[ \left(\Gamma_k^{(2)}+R_k \right)^{-1} k\, \partial_k R_k \right].
\ee
Here, $\Gamma_k^{(2)} = \frac{\delta^2\Gamma_k}{\delta \phi(x)\delta \phi(y)}$ denotes the second functional derivative. 
To extract information from Eq.~\eqref{eq:floweq}, we truncate $\Gamma_k$ as
\begin{equation}
	\begin{aligned}
		\Gamma_k 
		&= \int \frac{d^4p}{(2\pi)^4} f_{\rm Euclidean}(p^2) \,\phi(-p) \phi(p) \\
		&+ \frac{\lambda(k)}{4} \int \frac{d^4p_1}{(2\pi)^4}\frac{d^4p_2}{(2\pi)^4}\frac{d^4p_3}{(2\pi)^4}
		\,\phi(p_1)\phi(p_2)\phi(p_3)\phi(-p_1-p_2-p_3) \,
	\end{aligned}
\end{equation}
with $\lambda(k)$ being a scale-dependent quartic coupling. The argument of the last $\phi$ in the interaction term reflects momentum-conservation.
Plugging this truncation into Eq. \eqref{eq:floweq} and setting $\phi(x) = \bar{\phi} = \text{const.}$, we can project both sides onto the $\phi^4$-term, resulting in
\begin{align}
	k\, \partial_k \lambda(k) = \frac{1}{3!} \left[\,\frac{\partial^4}{\partial \bar{\phi}^4} \left( \frac{1}{2} \int \frac{d^4p}{(2\pi)^4} \frac{1}{f_{\rm Euclidean}(p^2)+ R_k(p^2) + 3 \lambda \bar{\phi}^2}k\, \partial_k R_k(p^2) \right) \right]_{\bar{\phi}=0}
\end{align}
The right-hand-side is both UV- and IR-finite: In the IR, for $p^2\rightarrow 0$, the presence of the regulator $R_k(p^2)$ in the propagator ensures finiteness. In the UV, for $p^2 \rightarrow \infty$, the scale-derivative of the regulator $k\partial_k R_k(p^2)$ in the denominator regularizes the momentum integral and imposes an effective cutoff of the momentum integration at $p^2 \gtrsim k^2$.\\
Setting $\phi$ on the right-hand-side to a constant is a choice that we can make in order to simplify the evaluation of the momentum integral, because the interaction term of interest, $\lambda/4 \,\phi^4$ is nonvanishing for this choice. In contrast, momentum-dependent interactions, which are in general also generated on the right-hand side, are discarded by this choice. 
Similarly, the 4th derivative with respect to $\bar{\phi}$ allow us to select the $\bar{\phi}^4$ term on the right-hand side, discarding higher-order interactions which are also generated.

\section{Assumptions underlying our analysis}\label{sec:assumptions}
Our analysis is based on a number of assumptions that enable us to link the scalar mass to the scale $\rho$. Here, we spell them out for clarity:
\begin{enumerate}
\item[a)] The Euclidean, causal-set inspired propagator defined in Eq.~\eqref{eq:euclprop} captures the main effects of causal-set-discreteness on the propagation of scalar fields.
\item[b)] The propagator of the theory is modified, but the interactions are not. There is in particular no momentum-dependent and potentially non-local dressing of the four-scalar interaction.
\item[c)] It makes sense to integrate out quantum fluctuations in the continuum QFT at momentum scales $p^2 > \sqrt{\rho}$, where $\rho^{-1/4} = \ell_k$ is the scale over which the discrete d'Alembertian is smeared, because $\ell_k \gg \ell$ must hold in order for the smearing to be effective in reducing fluctuations of the discrete d'Alembertian about the average value.
\item[d)] The local coarse graining at the heart of the functional RG continues to be suitable with the causal-set inspired nonlocal propagator.  
\end{enumerate}

\section{Bounds on the scalar mass in causal set quantum gravity}\label{sec:results}
The motivation for our work is the triviality bound on the Higgs mass, which provides an upper bound on the Higgs mass as a function of the scale up to which the theory is valid. We expect that the nonlocal modification of the propagator modifies such a mass bound. In turn, the connection between the Higgs mass and the scale where causal-set effects set in allows to bound this scale, given the measured value of the Higgs mass. To set the stage for this study, we first review how the triviality bound on the Higgs mass arises.

\subsection{Review of the triviality bound on the Higgs mass}
From the measurement of the Higgs mass, one can conclude about the scale of new physics. As a function of this scale, upper and lower bounds on the Higgs mass exist. In our work, the lower (stability) bound will not play a role. Instead, we focus on the upper (triviality) bound, see \cite{Maiani:1977cg,Cabibbo:1979ay,Dashen:1983ts,Callaway:1983zd,Beg:1983tu,Lindner:1985uk,Kuti:1987nr,Hambye:1996wb}. We first focus on a real scalar field $\phi$ with quartic coupling $\lambda$. In the symmetry-broken phase, where the scalar field acquires a vacuum expectation value 
\be
\langle \phi \rangle = v,
\ee
the mass of the excitations about this vacuum is given by
\be
m^2 = 3 \lambda\, v^2.
\ee
Allowed values of the mass $m$ therefore depend on allowed values of the quartic interaction $\lambda$. In turn, $\lambda$ is constrained to lie within a finite interval that is bounded from above by the triviality bound. The bound can already be motivated within one-loop perturbation theory, although its proof relies on nonperturbative techniques \cite{Maiani:1977cg,Cabibbo:1979ay,Dashen:1983ts,Callaway:1983zd,Beg:1983tu,Lindner:1985uk,Kuti:1987nr,Hambye:1996wb}.
At one loop, the beta function of the quartic coupling is given by
\be
\beta_{\lambda} = \beta_0\, \lambda^2,
\ee
where the one-loop coefficient $\beta_0>0$ depends on the normalization of the quartic interaction. For our choice of conventions, with $\lambda/4 \,\phi^4$, 
\be
\beta_0=\frac{9}{8\pi^2}.
\ee
The beta function can be integrated to yield
\be
\lambda(k)= \frac{\lambda_0}{1- \beta_0 \lambda_0 \ln \left(\frac{k}{k_0}\right)},\label{eq:lambdak}
\ee
where $\lambda_0=\lambda(k_0)$ with an infrared (IR) reference scale $k_0$ and the RG scale $k$. Because $\beta_0>0$, Eq.~\eqref{eq:lambdak} contains a pole at 
\be
k_{\rm Landau} =k_0\, e^{\frac{1}{\beta_0\, \lambda_0}}.\label{eq:Landaupole}
\ee
This pole signals the breakdown of perturbation theory, but, if we ignore the breakdown of the method\footnote{The resulting conclusions can be confirmed, e.g., with nonperturbative lattice simulations \cite{Luscher:1987ay,Luscher:1987ek} and other nonperturbative techniques \cite{Aizenman:1981du,Frohlich:1982tw}}, we can interpret Eq.~\eqref{eq:Landaupole} in physical terms. It implies that scalar field theory breaks down at a finite scale of new physics, $\Lambda_{\rm NP}$, which can be made larger by decreasing $\lambda_0$. Thus, the scalar field mass is bounded from above: the larger $\Lambda_{\rm NP}$, the smaller $\lambda_0$ and thus $m^2$ must be. More specifically, the mass is bounded from above by
\be
m_{\rm upper}^2 (k_{\rm Landau}) = 3 \lambda_0 (k_{\rm Landau})\, v^2,
\ee
where $\lambda_0(k_{\rm Landau})$ is the value of $\lambda_0$ that corresponds to a given choice of the Landau-pole scale.

Beyond perturbation theory, non-perturbative techniques, such as lattice and functional techniques, can be used. They confirm that one cannot extend scalar field theory up to arbitrarily high (momentum) scales, without requiring $\lambda_0$ to vanish. The triviality bound on the scalar mass therefore persists \cite{Aizenman:1981du,Frohlich:1982tw,Luscher:1987ay,Luscher:1987ek}.

For the Higgs field, $\beta_{\lambda}$ also depends on the $\text{SU}(2)_\text{L}$ and the $\text{U}(1)_\text{Y}$ gauge couplings, as well as the Yukawa couplings to the fermions (although the top quark Yukawa coupling is in practise the only non-negligible one). These additional couplings do not alter the existence of the triviality bound; they only alter the scale of new physics. Therefore, in order to make a first step towards deriving an upper bound on the Higgs mass in causal set quantum gravity, we discard all additional degrees of freedom of the Standard Model, and focus on a real scalar field.

\subsection{Triviality bound on scalar $\phi^4$ theory with the causal-set inspired propagator}
\begin{figure}[!t]
\includegraphics[width=0.6\linewidth]{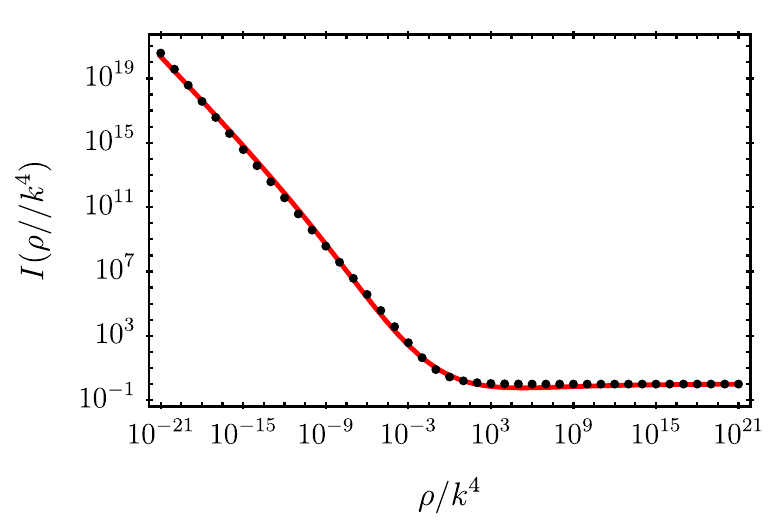}
\caption{\label{fig:loopintegral} We show the numerical data to the result of the loop integral, cf.~Eq.~\eqref{eq:loopint} (black dots) as well as the fit, cf.~Eq.~\eqref{eq:loopfit} (red line).}
\end{figure}

The beta function for $\lambda$ depends on a loop integral over quantum fluctuations, obtained from the flow equation \eqref{eq:floweq} and we can write it as
\be
\beta_{\lambda}=k\partial_k\lambda(k) = \frac{9}{8\pi^2}\lambda^2 I(\rho/k^4),\label{eq:betalambdaCS}
\ee
where $I(\rho/k^4)$ depends on the nonlocality scale.
In terms of $r(p^2/k^2) = R_k(p^2)/k^2$ and $y=p^2/k^2$, we can write
\begin{equation}
	I(\rho/k^4) = \int_0^\infty dy\, \frac{2y\,(r(y)- y r'(y))}{\left(k^{-2} f_\textmd{Euclidean}\left(k^2 \,y \right)+ r(y) \right)^3} \,.
\end{equation}
This loop integral over quantum fluctuations decomposes into three parts: The first is the UV part, in which the propagator takes the asymptotic form Eq.~\eqref{eq:UVlimit}, the second is the intermediate part, in which $p^2 \approx \sqrt{\rho}$ and which can only be performed numerically, and the third is the IR part, in which the propagator takes the standard form. The IR and UV parts, where the propagator assumes a simple form, can be performed analytically, the intermediate part numerically. 
 We can interpolate the final result in terms of the four-parameter function
\begin{equation}
	I(\rho/k^4) = \left(\frac{\rho}{\rho_0\, k^4} \right)^{\frac{a_1}{1+a_2 \left(\frac{\rho}{\rho_0\, k^4}\right)^{\alpha}}}\label{eq:loopfit} \,,
\end{equation}
with fitting parameters
\be
a_1=-0.89,\quad a_2=3.96,\quad \alpha=0.14, \quad \rho_0=83.30.
\ee
We show both the numerical result as well as the fit in Fig.~\ref{fig:loopintegral}.

Based on this result, the beta function Eq.~\eqref{eq:betalambdaCS} can be integrated to yield the running coupling
\be
\lambda(k) = \lambda_0 \left( 1 - \frac{9\, \lambda_0}{8\pi^2} \int_{k_0}^{k} dk'\, \frac{I(\rho/{k'}^4)}{k'}\right)^{-1} \,.
\ee
The right-hand side needs to be evaluated numerically. We can rewrite it in terms of a new function $F(k/\rho^{1/4})$, which can be evaluated numerically. With
\begin{align}\label{eq:IntF}
	F(z) = \int_1^{z} d\tilde{k}\,\frac{I(\tilde{k}^{-4})}{\tilde{k}} \,,
\end{align}
we can write 
\begin{align}\label{eq:lambdasol}
	\lambda(k) = \lambda_0 \left( 1 - \frac{9\, \lambda_0}{8\pi^2} \,
	\Big[ F\big( k/\rho^{1/4}\big) - F\big( k_0/\rho^{1/4}\big) \Big]\right)^{-1} \,.
\end{align}

\begin{figure}[!t]
	\begin{center}
		\includegraphics[width=0.6\linewidth]{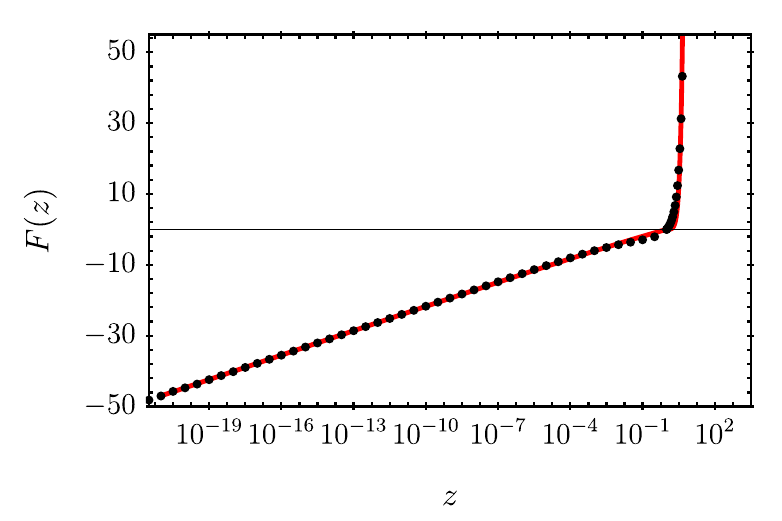}
		\caption{Numerical data (black dots) and corresponding fit (red line) of the function $F(z)$.}
		\label{fig:FLitim}
	\end{center}
\end{figure}

For $F(z)$, we find a three-parameter fit (cf.~Fig.~\ref{fig:FLitim})
\begin{align}\label{eq:IntFfit}
	F(z) = \big( \mathcal{A} \left(z^a-1\right)+\log (z) \big)\, \theta(1-z)\, + z^b \,\theta(z-1) \,.
\end{align}
with the fitting parameters
\begin{align}
	\{ \mathcal{A}, a , b  \}_\textmd{Fit} = \{ -1.363 ,\,0.126 , \,3.161\} \,.
\end{align}

Plugging \eqref{eq:IntFfit} into \eqref{eq:lambdasol}, we find
\begin{equation}
	\begin{aligned}
		\lambda(k) &= 
		\lambda_0 \left[ 1 - \frac{9\lambda_0}{8\pi^2} \left( \log \big(k/k_0 \big)
		+ \mathcal{A} \, \rho^{-a/4} \left( k^a - k_0^a \right) \right) \right]^{-1} \theta\big(1-k/\rho^{1/4}\big)  \\
		&+ \lambda_0  \left[ 1 - \frac{9\lambda_0}{8\pi^2} \left( \log \big(\rho^{1/4}/k_0 \big)
		+ \mathcal{A} \, \rho^{-a/4} \left( \rho^{a/4} - k_0^a \right) \right) 
		+ \big(k/\rho^{1/4}\big)^b\right]^{-1} \theta\big(k/\rho^{1/4}-1\big) \,,\label{eq:runninglambda}
	\end{aligned}
\end{equation}
where we have used $k_0<\rho^{1/4}$.
As a consistency check, we confirm that for $\rho \to \infty$, we recover 
\begin{align}
	\lambda_{\rho\to\infty}(k)=\lambda_0 \left[ 1 - \frac{9\lambda_0}{8\pi^2} \log \big(k/k_0 \big)  \right]^{-1} \,,
\end{align}
which agrees with the usual result for the running of the quartic coupling $\lambda$ obtained with the standard d'Alembertian operator.

We can now analyze the physical implications of our result, Eq.~\eqref{eq:runninglambda}. We first note that, due to existence of a fixed external scale, $\sqrt{\rho}$, there are two regimes in the running of $\lambda$: For $k^2 < \sqrt{\rho}$, i.e., in the IR, we recover the logarithmic scale dependence of $\lambda$ with a small correction $\sim (k/\rho^{1/4})^a = (k/\rho^{1/4})^{0.126}$. For $k^2 >\sqrt{\rho}$, in the UV, the scale dependence is strongly modified and we see a dependence on $(k/\rho^{1/4})^b=(k/\rho^{1/4})^{3.161}$. 

We now follow the RG flow in the opposite direction, i.e., in the direction of increasing $k$. As soon as $k$ becomes comparable to $\rho^{1/4}$, the modified form of the propagator results in a strongly accelerated flow, which speeds up the onset of the Landau pole. We thus expect to encounter a Landau pole close to the nonlocality scale. 

There is a second perspective on this result, namely that the nonlocality dampens the interaction. This becomes clear if one follows the RG flow in its proper direction, from microscopic to macroscopic scales (i.e., from large to small $k$): then, even large values of $\lambda$ are very quickly driven towards small values; the nonlocality thus accelerates the approach of $\lambda$ to zero.

Both perspectives imply that the Landau-pole problem persists: Seen from the IR, the nonlocality speeds up the onset of the Landau pole. Seen from the UV, the nonlocality speeds up the decrease of the coupling towards the IR; thus, a very large $\lambda$ has to be chosen in the UV in order for $\lambda$ to differ significantly from 0 in the IR. This second way of viewing the result is in spirit closely related to the triviality problem. We caution that to confirm the triviality problem, a fully nonperturbative study would be necessary. The functional RG is in principle a well-suited tool to do so. However, our present truncation, where we only account for the quartic interaction, and discard both higher-order interactions as well as momentum-dependences in the interaction, is not well-suited to explore the nonperturbative regime.

\begin{figure}[!t]
	\begin{center}
		\includegraphics[width=0.6\linewidth]{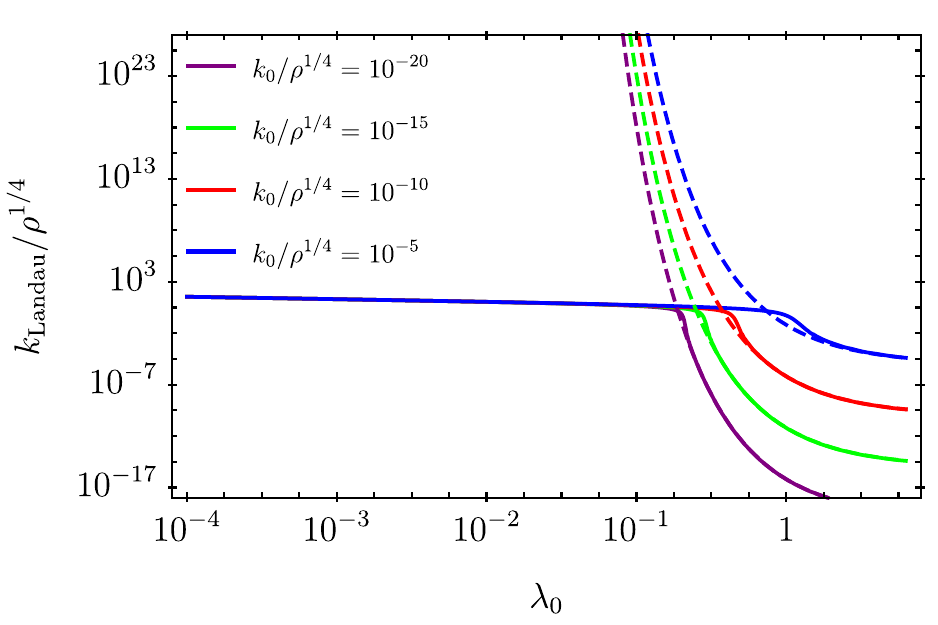}
		\caption{We show the Landau pole scale $k_\textmd{Landau}$ as a function of $\lambda_0$ for various choices of the ratio $k_0/\rho^{1/4}$. 
		The full lines correspond to the results obtained in our CS-inspired setting. The dashed lines correspond to $k_\textmd{Landau} = k_0 \exp\left( \frac{8\pi^2}{9\lambda_0}\right)$ obtained with the standard $\lambda \phi^4$-model.}
		\label{fig:kLandau}
	\end{center}
\end{figure}

We now quantify the scale of the Landau pole, $k_{\rm Landau}$ and in particular its relation to the nonlocality scale. $k_{\rm Landau}$ follows by solving 
\begin{align}\label{eq:LandauPole}
	1 - \frac{9\, \lambda_0}{8\pi^2} \,
	\Big[ F\big( k_\textmd{Landau}/\rho^{1/4}\big) - F\big( k_0/\rho^{1/4}\big) \Big] = 0\,.
\end{align}
Assuming that $F$ has an inverse, we can write the formal solution
\begin{align}
	k_\textmd{Landau}=\rho^{1/4} \,F^{-1}\left( \frac{8\pi^2}{9\lambda_0} + F\big( k_0/\rho^{1/4}\big)\right) \,.\label{eq:Landaures}
\end{align}
For our purposes, it is not relevant whether $F^{-1}$ exists on the entire domain of definition of $F(z)$; we only care about the numerical inverse for a finite range of arguments, in which, from Fig.~\ref{fig:FLitim}, one can see that the inverse exists. We evaluate the inverse numerically from our numerical data for $F(z)$ and thereby determine the Landau-pole scale, cf.~Fig.~\ref{fig:kLandau}.

For small values of $\lambda_0$, we obtain a result that does not depend on the choice of $k_0$, as long as $k_0 \ll \rho^{1/4}$. This follows directly from the fact that if $\lambda_0$ is small enough, the first term in the argument of $F^{-1}$ dominates, irrespective of the choice of $k_0/\rho^{1/4}$.
In this regime we obtain a universal limit, namely that the Landau-pole scale is roughly equal to the nonlocality scale. This result confirms our expectation based on the properties of the propagator: because UV modes are not as usual suppressed by a $1/p^2$ propagator, and instead strongly enhanced to a constant propagator, their effect on the running coupling is compounded and the Landau pole sets in much earlier than in standard QFT.

For larger values of $\lambda_0$, the Landau-pole scale depends on the choice of $k_0$. In this regime, because the initial condition for $\lambda$ is already close to the nonperturbative regime, the Landau-pole is approached close to $k_0$. Thus, the Landau-pole scale can occur significantly below the nonlocality scale in this case.  In this regime, we find agreement with standard QFT, because even a local propagator suffices to drive the system into a Landau pole if the initial condition for $\lambda$ is chosen in the nonperturbative regime.

\begin{figure}[!t]
\includegraphics[width=1\linewidth]{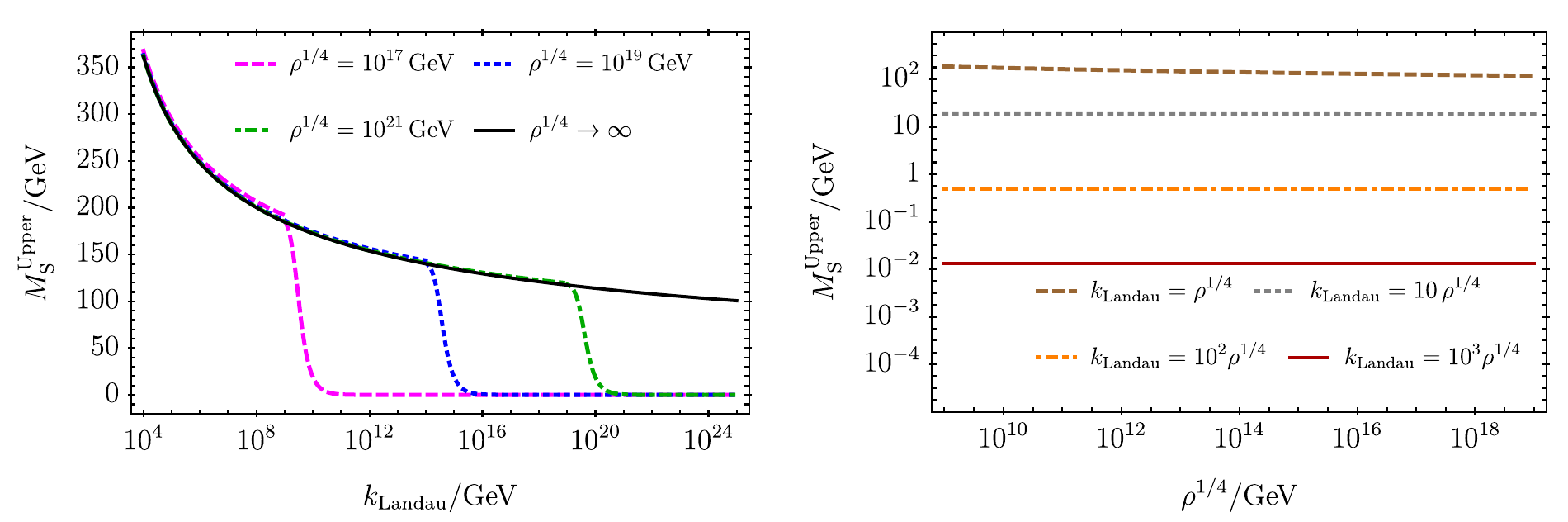}
\caption{\label{fig:massbounds} Left panel: We show the upper bound on the scalar mass as a function of the Landau-pole scale for different values of the nonlocality scale.
Right panel: We show the upper bound on the scalar mass as a function of the nonlocality scale for different values of the ratio $k_\textmd{Landau}/\rho^{1/4}$.}
\end{figure}

Finally, we can translate our results into an upper bound on the scalar mass as a function of the Landau-pole scale. 
We demand that the theory is valid up to the Landau pole scale $k_\textmd{Landau}$ and thus obtain the bound
\begin{align}
	 \lambda_0  < \frac{8\pi^2/9}{F\big( k_\textmd{Landau}/\rho^{1/4}\big) - F\big( k_0/\rho^{1/4}\big)}\,.
\end{align}
This translates into an upper bound on the scalar mass, evaluated at $k_0$
\begin{align}
	M_\text{S}(k_0) \lesssim \left( \frac{8\pi^2  v_\textmd{EW}^2/9}{F\big( k_\textmd{Landau}/\rho^{1/4}\big) - F\big( k_0/\rho^{1/4}\big)} \right)^{\!1/2} = M_\text{S}^\textmd{Upper}\,.
\end{align}
For our numerical evaluation, we choose $v_{\rm EW} = 246 \, \rm GeV$ and $k_0=173\, \rm GeV$, as relevant for the Higgs particle in the Standard Model, cf.~Fig.~\ref{fig:massbounds}. We see that as soon as the Landau-pole scale lies close to the nonlocality scale, the window of allowed scalar masses closes. Thus, if we extrapolate our result to the Standard Model, the nonlocality scale may not lie significantly below the Planck scale, because otherwise new physics is needed in a regime where by assumption the standard QFT still holds. One would then have to embed the Standard Model into a UV complete (i.e., asymptotically free or safe) setting and hope that the presence of the nonlocality does not destroy the UV completion.

We conclude that a significant separation between the nonlocality scale and the discreteness scale, as is required in order to dampen the fluctuations in the d'Alembertian, appears difficult to reconcile with the absence of Landau poles. Instead, the nonlocality appears to speed up the breakdown of perturbation theory (and, if the results persist nonperturbatively, of the theory as a whole), and requires new physics close to the nonlocality scale. The new physics may potentially ``just'' be new fields, but it is unclear whether one can actually avoid similarly problematic features just by adding new fields and interactions. It may therefore be the case that the nonlocality must be confined to the discreteness scale.

\section{Conclusions and outlook}\label{sec:conclusions}
In the spirit of the ``matter matters" paradigm, we set out to explore the implications of causal set discreteness for the upper bound on the Higgs mass. We take the first step in this exploration by focusing on a scalar field with quartic interaction, and neglect both the gauge symmetries as well as the additional fields of the Standard Model. We use the causal set d'Alembertian, for which an expression in Fourier space is available, once the d'Alembertian is averaged over sprinklings into Minkowski spacetime. In this averaging, fluctuations about the mean grow with sprinkling density and  it has therefore been proposed that the expression should be smeared over a distance scale $\rho^{-1/4}$ which is significantly larger than the discreteness scale. The resulting d'Alembertian deviates from the standard d'Alembertian at this distance scale, where it becomes nonlocal; thus the scale is referred to as the nonlocality scale. 

We find that the running of the quartic scalar coupling is modified at the nonlocality scale. Because the nonlocality enhances quantum fluctuations, it accelerates the running of the coupling. Therefore, the Landau pole, which occurs in the quartic scalar interaction already in standard QFT, is still present, but shifted to significantly lower scales and occurs close to the nonlocality scale.

We assume that, just as in standard QFT, the Landau pole signals not just the breakdown of perturbation theory, but the breakdown of the theory itself and the need for new physics. This new physics could in principle be physics beyond the Standard Model, i.e., new matter fields. It is, however, unclear, whether the enhancement of quantum fluctuations with momenta beyond the nonlocality scale enables the extension of a QFT significantly beyond the nonlocality scale. We thus tentatively interpret the nonlocality scale as the scale at which continuum QFT may break down. This implies that the separation between nonlocality scale and discreteness scale, required to suppress fluctuations of the d'Alembertian, may not be consistent.

It is also possible to associate $\rho^{-1/4}$ directly with the discreteness scale, in which case the underlying discrete d'Alembertian is not ``smeared". This interpretation of $\rho$ is favored by our results, because the Landau pole can then be viewed as signalling the breakdown of continuum quantum field theory and the onset of discreteness. It is then selfconsistent that the Landau pole occurs roughly at $\rho^{-1/4}$.

\subsection{Outlook on methodological developments}
The Lorentzian causal set propagator, with the constant contribution subtracted, shows an improved UV behavior, which can regularize loop integrals. This paves the way for an upgrade of the present study to Lorentzian signature, combining a Callan-Symanzik IR cutoff with the Lorentzian causal-set propagator.

In addition, the propagator may be used in the context of asymptotically safe quantum gravity, where studies of Lorentzian signature are technically challenging \cite{Manrique:2011jc,Bonanno:2021squ,Fehre:2021eob} and where a UV regularization compatible with Lorentzian signature is required.

\subsection{Outlook on physics implications}
Ours is not the first work which aims at relating the mass of a scalar field to the scale of quantum gravity. In \cite{Shaposhnikov:2009pv}, it was previously proposed that asymptotically safe quantum gravity gives rise to a prediction of the Higgs mass.  There is therefore the perspective that distinct quantum-gravity approaches give rise to distinct predictions of the Higgs mass, or distinct upper bounds on it. These are of course subject to assumptions, such as, e.g., the absence of new fields that couple to the Higgs boson. Under these assumptions, such distinct predictions and/or bounds would provide an opportunity to select among the different contenders for a quantum theory of gravity. 

\subsection{Outlook on quantum gravity fluctuations}
In causal set quantum gravity, quantum gravity fluctuations are fluctuations of the causal set. These are not directly related to fluctuations in the metric, or fluctuations of the graviton. To add the impact of graviton-fluctuations to our study, one may therefore proceed in the following way: First, we assume that graviton fluctuations become important at scales at which the non-local d'Alembertian encodes the propagation of scalar fields. Second, we assume that the d'Alembertian also encodes the propagation of gravitons.\footnote{Their propagator is then the usual transverse traceless projector, multiplied with the inverse d'Alembertian instead of the usual factor of one over momentum squared.} Third, we assume that the coupling of gravitons to the scalar is given by the usual vertices from minimal coupling. Under those assumptions, we can calculate the impact of gravitons on the beta function of the scalar quartic coupling. We obtain a contribution that is linear in the scalar quartic coupling and is positive at all scales\footnote{At momentum scales below $\rho^{1/4}$, this corresponds to the standard contribution that has, e.g., been evaluated in the context of asymptotically safe quantum gravity \cite{Narain:2009fy}.} Thus, we conclude that, under the three assumptions above, graviton fluctuations cannot prevent the onset of a Landau pole at the nonlocality scale.

\begin{acknowledgments}   
This work is supported by a research grant (29405) from VILLUM fonden. A.~E.~acknowledges support from the Perimeter Institute for Theoretical Physics during the final stages of this work. Research at Perimeter Institute is supported in part by the Government of Canada through the Department of Innovation, Science and Economic Development and by the Province of Ontario through the Ministry of Colleges and Universities.
 \end{acknowledgments}

\bibliography{refs}
\end{document}